\documentclass{article}

\usepackage{microtype}
\usepackage{graphicx}
\usepackage{subcaption}
\usepackage{booktabs} 

\usepackage{hyperref}

\usepackage[accepted]{icml2026}

\usepackage{amsmath}
\usepackage{amssymb}
\usepackage{mathtools}
\usepackage{amsthm}
\usepackage{graphicx}
\usepackage{booktabs}
\usepackage{multirow}
\usepackage{amsmath}
\usepackage{subcaption}
\usepackage{listings}
\usepackage{float}
\usepackage{makecell}

\usepackage[capitalize,noabbrev]{cleveref}

\theoremstyle{plain}

\theoremstyle{definition}

\theoremstyle{remark}

\usepackage[textsize=tiny]{todonotes}

\icmltitlerunning{Succeeding at Scale}

\begin{document}

\twocolumn[
  \icmltitle{Succeeding at Scale: Enterprise Retrieval Benchmark Construction and Index-Preserving Query Adaptation for Multi-Tenant Search}



  \icmlsetsymbol{equal}{*}

\begin{icmlauthorlist}
  \icmlauthor{Prateek Jain}{equal,devrev-austin}
  \icmlauthor{Shabari S Nair}{equal,utexas}
  \icmlauthor{Ritesh Goru}{devrev-austin}
  \icmlauthor{Prakhar Agarwal}{devrev-bengaluru}
  \icmlauthor{Ajay Yadav}{utexas}
  \icmlauthor{Yoga Sri Varshan Varadharajan}{utexas}
  \icmlauthor{Constantine Caramanis}{utexas}
\end{icmlauthorlist}

\icmlaffiliation{devrev-austin}{DevRev, Austin, USA}
\icmlaffiliation{utexas}{The University of Texas at Austin, Austin, USA}
\icmlaffiliation{devrev-bengaluru}{DevRev, Bengaluru, India}

\icmlcorrespondingauthor{Prateek Jain}{prateek.jain@devrev.ai}
\icmlcorrespondingauthor{Shabari S Nair}{shabarisnair@utexas.edu}

\icmlkeywords{Machine Learning, Information Retrieval, Dense Retrieval, Multi-Tenant Search, Query Adaptation, Low-Rank Adaptation, LoRA, Parameter-Efficient Fine-Tuning}

\vskip 0.3in
]



\printAffiliationsAndNotice{\icmlEqualContribution}

\begin{abstract}
  Large-scale multi-tenant retrieval systems generate extensive query logs but lack curated relevance labels for effective domain adaptation, resulting in substantial underutilized “dark data.” This challenge is compounded by the high cost of model updates, as jointly fine-tuning query and document encoders requires full corpus re-indexing, which is impractical in multi-tenant settings with thousands of isolated indices. We introduce DevRev-Search, a passage retrieval benchmark for technical customer support built via a fully automated pipeline. Candidate generation uses fusion across diverse sparse and dense retrievers, followed by an LLM-as-a-Judge for consistency filtering and relevance labeling. We further study and systematically evaluate index-preserving query-only adaptation strategies that fine-tune only the query-encoder while keeping the document indices fixed. Experiments on DevRev-Search, SciFact, and FiQA-2018 show that parameter-efficient fine-tuning of the query encoder delivers a remarkable quality–efficiency trade-off, enabling scalable and practical enterprise multi-tenant retrieval.
  
\end{abstract}

\section{Introduction}

The transition from lexical matching methods such as BM25 \cite{robertson2009bm25} to dense neural retrieval has revolutionized information discovery \cite{karpukhin-etal-2020-dense}. However, deploying bi-encoder architectures in multi-tenant enterprise environments presents two major challenges. First, the \textbf{Data Scarcity Bottleneck}: enterprise tenants possess proprietary "dark data" where relevance labels are unavailable, and standard benchmarks like BEIR \cite{thakur2021beir} fail to capture the noisy, heterogeneous nature of these domains. Second, the \textbf{Adaptation Latency Bottleneck}: symmetric fine-tuning of both encoders incurs a massive "Re-indexing Tax," as any document encoder update requires regenerating embeddings for the entire corpus, which is computationally prohibitive for platforms hosting thousands of tenants.

To address these challenges, we present a unified methodology for scalable dataset construction and efficient model adaptation. Our main contributions are as follows:
\begin{enumerate}
    \item \textbf{DevRev-Search Benchmark \footnote{The DevRev-Search dataset is made available at \href{https://huggingface.co/datasets/devrev/search}{https://huggingface.co/datasets/devrev/search}.}}: We construct a scalable pipeline that synthesizes training data without human annotation by pooling candidates from diverse retrievers and applying LLM-as-a-Judge filtering \cite{dai2023promptagator, rahmani2024llmjudge}.
    \item \textbf{Zero-Reindexing Adaptation}: We propose a new approach to fine-tuning dense retrieval models by only adapting the query encoder while keeping the document encoder (and indices) fixed.
    \item \textbf{Parameter-Efficient Query Adaptation}: We comprehensively evaluate techniques such as LoRA \cite{hu2022lora}, embedding transformations via linear and feed-forward projections, and selective fine-tuning of transformer layers, as efficient query-only fine-tuning strategies. We also conduct ablations over LoRA ranks, modules, and the number of unfrozen layers to find the best possible configurations.
\end{enumerate}

Through extensive evaluation on DevRev-Search, SciFact \cite{wadden2022scifact}, and FiQA-2018 \cite{fiqa2018}, we demonstrate that Parameter-Efficient query-only fine-tuning  achieves robust domain adaptation, thus enabling tenant-specific models for multiple tenants at a fraction of the cost of full bi-encoder fine-tuning. In summary, we develop a unified framework for enterprise multi-tenant retrieval that combines a new scalable benchmark with a new, cost-effective domain adaptation approach based on query-only adaptation for dense retrieval. To the best of our knowledge, such an approach has not been explored in prior work, particularly in multi-tenant settings.



\section{Related Work}

\textbf{Synthetic Data for Retrieval:} Recent work leverages LLMs for synthetic data generation \cite{bonifacio2022inparsdataaugmentationinformation, wang-etal-2022-gpl, dai2023promptagator}. Inspired by the success of combining multiple retrievers \cite{cormack2009rrf}, we adopt \textbf{fusion-based candidate generation}, aggregating diverse retrievers to reduce single-model bias and improve coverage. \textbf{LLM-as-a-Judge:} While LLMs show promise as relevance assessors \cite{rahmani2024llmjudge}, bias concerns remain \cite{soboroff2025dontusellms}. We therefore use LLMs primarily for filtering pooled candidates, identifying positives rather than generating them, and validate a subset with human annotators. \textbf{Index-Preserving Adaptation:} Standard symmetric fine-tuning \cite{karpukhin-etal-2020-dense} incurs a prohibitive "Re-indexing Tax." Prior index-preserving approaches focus on pseudo-relevance feedback \cite{yu2021anceprf} or asymmetric tuning \cite{wang2023queryencoderdistillationembedding}. We propose \textbf{Query-Only Adaptation} and study it as an index-preserving deployment strategy in multi-tenant retrieval environments. \textbf{Parameter-Efficient Fine-Tuning (PEFT):} Methods such as Adapters \cite{houlsby2019adapters}, LoRA \cite{hu2022lora}, embedding projections \cite{yoon-etal-2024-search}, and partial layer unfreezing \cite{lee2022surgical} have proven effective in retrieval \cite{litschko2022parameterefficientneuralrerankingcrosslingual}. Building on findings about intrinsic dimensionality \cite{aghajanyan2020intrinsicdimensionalityexplainseffectiveness}, we show that applying PEFT to the query encoder achieves both strong performance and efficiency for multi-tenant serving.

\section{Dataset Generation}

\begin{table*}[t]
\caption{\label{tab:main_table}Comparison of Recall@10 and NDCG@10 in retrieval effectiveness for baseline untrained model (Base) with Query-Document (QD), and Query-Only (Q) finetuning}
\begin{center}
\begin{tabular}{llcccccc}
\toprule
\multirow{2}{*}{Model} & \multirow{2}{*}{Variant} 
& \multicolumn{2}{c}{DevRev-Search} 
& \multicolumn{2}{c}{SciFact} 
& \multicolumn{2}{c}{FiQA} \\
\cmidrule(lr){3-4} \cmidrule(lr){5-6} \cmidrule(lr){7-8}
& & recall@10 & ndcg@10 & recall@10 & ndcg@10 & recall@10 & ndcg@10 \\
\midrule
\multirow{3}{*}{arctic-l-v2}
& Base & 0.256 & 0.304 & 0.819 & 0.691 & 0.533 & 0.459 \\
& QD   & \textbf{0.314} & \textbf{0.362} & \textbf{0.869} & \textbf{0.748} & \textbf{0.578} & \textbf{0.505} \\
& Q    & 0.296 & 0.343 & 0.854 & 0.727 & 0.554 & 0.478 \\
\midrule
\multirow{3}{*}{qwen3-4b}
& Base & 0.219 & 0.264 & 0.903 & 0.769 & 0.649 & 0.571 \\
& QD   & 0.325 & \textbf{0.396} & 0.949 & \textbf{0.875} & \textbf{0.712} & \textbf{0.628} \\
& Q    & \textbf{0.327} & 0.370 & \textbf{0.953} & 0.830 & 0.688 & 0.602 \\
\bottomrule
\end{tabular}
\end{center}
\end{table*}

Recent retrieval gains are driven by high-quality, domain-specific data, yet publicly available enterprise search datasets remain scarce. Existing benchmarks like BEIR \cite{thakur2021beir} do not capture the semi-structured, heterogeneous nature of enterprise data, including support tickets, issue trackers, and internal documentation. We seek to address this gap by providing a high-fidelity benchmark for enterprise-specific retrieval. 

Manual annotation is costly and inherently low-recall, as annotators cannot exhaustively review large corpora, resulting in systematic false negatives. To address this, we introduce a scalable, automated dataset construction pipeline that maximizes candidate coverage through multi-stage retrieval while maintaining high precision via an LLM-as-a-judge.

\subsection{Query Collection and Cleaning}

We collected customer queries from production agent interactions as our source of real-world question data. However, raw customer queries often contain noise, including test queries, code snippets, and malformed inputs that are not legitimate natural language questions. To ensure dataset quality, we implemented a multi-stage filtering process: (1) \textbf{Length filtering}: Removing queries with word counts in the bottom and top 25\% percentiles. (2) \textbf{Language detection}: Retaining only English queries. (3) \textbf{Deduplication}: Removing exact duplicate queries. (4) \textbf{Clustering-based diversity}: Selecting representative samples from clusters to ensure diversity and avoid semantic repetition. For further details on the above steps, please refer to Appendix \ref{app:query_cleaning}

\subsection{Document Segmentation and Semantic Granularity}

Enterprise documents are typically long and sparsely relevant to specific queries, which poses challenges for dense retrieval due to encoder token limits and the limited expressive capacity of fixed-size embeddings. Encoding entire documents often results in diluted representations that obscure fine-grained relevance. To address this, we collected all public documentation of DevRev and applied Recursive Character Splitting \footnote{Documentation for LangChain's Recursive Text Splitter is available at \href{https://docs.langchain.com/oss/python/integrations/splitters/recursive_text_splitter}{the official LangChain documentation}.}. Documents are segmented into chunks of up to 500 characters with no overlap, maximizing semantic coverage while preserving precision. The recursive strategy favors natural structural boundaries (e.g., paragraphs, sentences, then spaces), ensuring each fragment is both embedding-friendly and semantically self-contained.

\subsection{Multi-Retriever Annotation}

To create high-quality query-document pairs, we employed an ensemble retrieval approach designed to maximize recall while maintaining precision.

\textbf{Retrieval Ensemble:} We applied an ensemble of seven diverse models: six dense retrievers (gemini-embedding-001 \cite{gemini_embed}, text-embedding-3-large \cite{openai_embed}, embed-english-v3 \cite{cohere_embed}, Qwen-3-Embedding-8B \cite{qwen3embedding}, GTE-Qwen2-7B-Instruct \cite{li2023towards}, SFR-Embedding-Mistral \cite{SFRAIResearch2024}) and one lexical retriever - BM25 \cite{robertson2009bm25}, each returning the top 60 document chunks.

\textbf{Union-based Aggregation:} We computed the union of results from all retrievers to create a comprehensive candidate set of potentially relevant chunks. This ensures that chunks retrieved by any of the seven models are included in the candidate set, maximizing coverage and recall across different retrieval paradigms yielding $\geq 60$ and $\leq 420$ unique candidate chunks per query across all models.

\textbf{LLM-based Filtering:} While ensembling improves recall, it also introduces false positives. To improve precision, we applied LLM-based filtering to the fused candidates using a prompt (Appendix~\ref{app:llm_filter_prompt}) that retains only chunks genuinely relevant to each query, removing lexically or semantically similar chunks lacking substantive answer content.


\textbf{Quality Validation:} To verify the reliability of our automated annotation process, we randomly sampled $10\%$ queries and manually validated the final annotations to have a precision of 92\%, which confirmed its effectiveness.

\paragraph{Retriever Contribution Analysis:} Using the pipeline-generated chunks as ground truth, we evaluated individual retriever performance. A maximum recall of 82.48 for gemini-embedding-001 (Table~\ref{tab:ind_cont}, Appendix~\ref{app:retriever_contrib}) shows that single retrievers miss many relevant documents. We also performed leave-one-out ablations by aggregating candidates from the remaining six retrievers. As shown in Table~\ref{tab:red_cont} (Appendix~\ref{app:retriever_contrib}), recall varies from 93.25 to 97.13, indicating that each retriever contributes unique candidates and that combining dense and lexical retrieval improves coverage.



Our pipeline yields a high-quality dataset of 383 queries and 65.2K chunks, termed DevRev-Search, split into 291 training and 92 test queries. The dataset has high relevance density, averaging 13.61 golden chunks per query, with further details provided in Appendix~\ref{app:dataset_statictics}.

\section{Query-Only Adaptation}
\label{sec:results}

\begin{table*}[t]
\caption{\label{tab:linear_head}Comparison of retrieval effectiveness for various parameter-efficient methods with full fine-tuning for query-only adaptation (Here *LoRA and 8-Tr refer to the best LoRA configuration and fine-tuning only 8 transformer layers at the top of the base model, respectively)}
\begin{center}
\begin{tabular}{llcccccc}
\toprule
\multirow{2}{*}{Model} & \multirow{2}{*}{Type} 
& \multicolumn{2}{c}{DevRev-Search} 
& \multicolumn{2}{c}{SciFact} 
& \multicolumn{2}{c}{FiQA} \\
\cmidrule(lr){3-4} \cmidrule(lr){5-6} \cmidrule(lr){7-8}
& & recall@10 & ndcg@10 & recall@10 & ndcg@10 & recall@10 & ndcg@10 \\
\midrule
\multirow{3}{*}{arctic-l-v2}
& Linear  & 0.271 & 0.314 & 0.849 & 0.722 & 0.543 & 0.470 \\
& FFN   & 0.281 & 0.318 & 0.880 & 0.743 & 0.534 & 0.464 \\
& *LoRA   & \textbf{0.309} & 0.342 & \textbf{0.915} & \textbf{0.827} & \textbf{0.555} & \textbf{0.478} \\
& 8-Tr   & 0.273 & 0.317 & 0.849 & 0.718 & 0.547 & 0.476 \\
& Full & 0.296 & \textbf{0.343} & 0.854 & 0.727 & 0.554 & \textbf{0.478} \\
\midrule
\multirow{3}{*}{qwen3-4b}
& Linear    & 0.331 & 0.358 & 0.935 & 0.815 & 0.666 & 0.583 \\
& FFN   & 0.340 & 0.362 & 0.946 & \textbf{0.861} & 0.660 & 0.577 \\
& *LoRA   & \textbf{0.355} & \textbf{0.399} & 0.950 & 0.844 & \textbf{0.694} & \textbf{0.607} \\
& 8-Tr   & 0.312 & 0.356 & 0.932 & 0.802 & 0.673 & 0.590 \\
& Full & 0.327 & 0.370 & \textbf{0.953} & 0.830 & 0.688 & 0.602 \\

\bottomrule
\end{tabular}

\end{center}
\end{table*}

\begin{figure*}[t]
    \centering
    \begin{subfigure}[b]{0.32\textwidth}
        \centering
        \includegraphics[width=\textwidth]{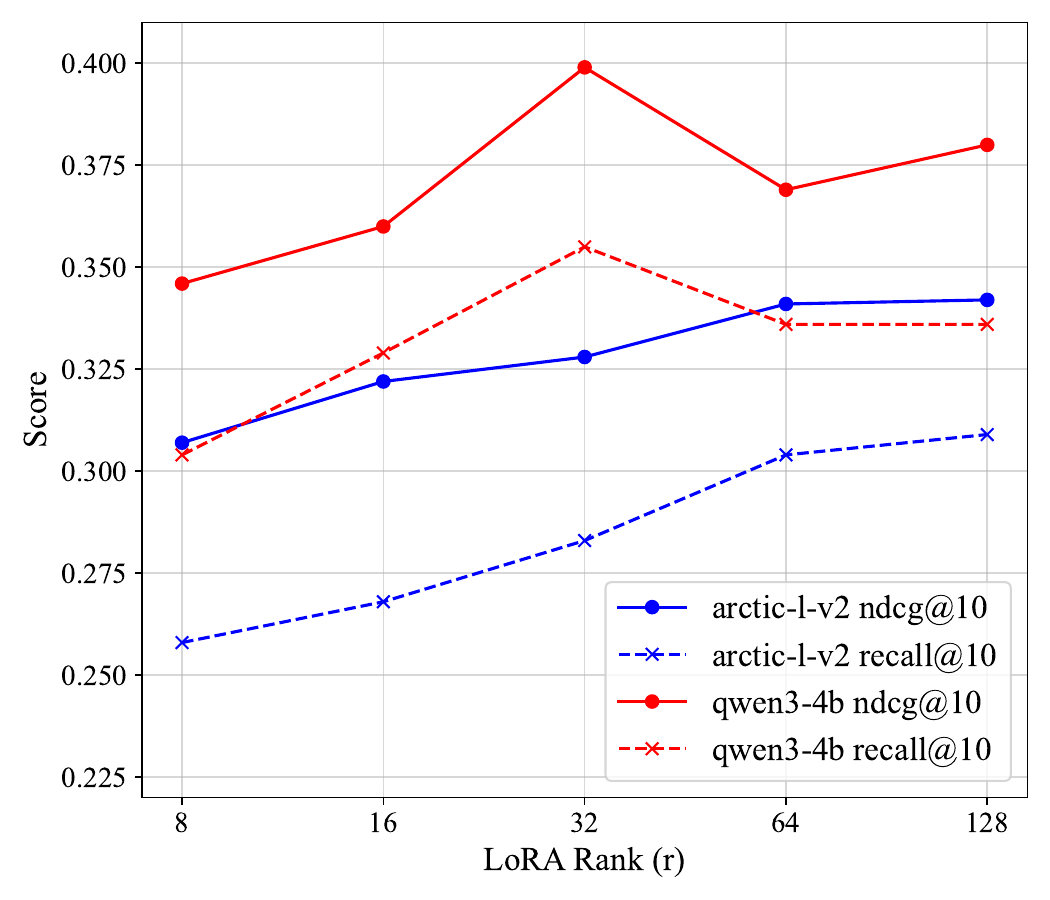}
        \caption{}
        \label{fig:a}
    \end{subfigure}
    \hfill
    \begin{subfigure}[b]{0.32\textwidth}
        \centering
        \includegraphics[width=\textwidth]{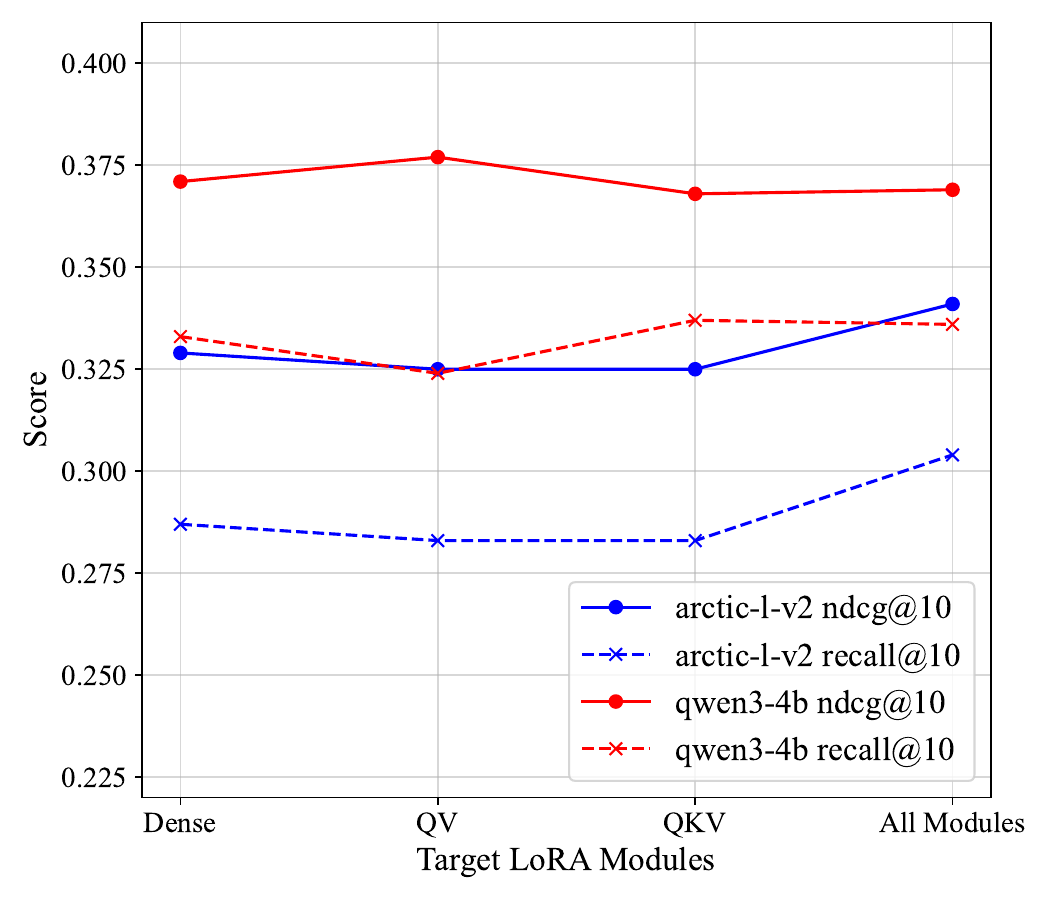}
        \caption{}
        \label{fig:b}
    \end{subfigure}
    \hfill
    \begin{subfigure}[b]{0.32\textwidth}
        \centering
        \includegraphics[width=\textwidth]{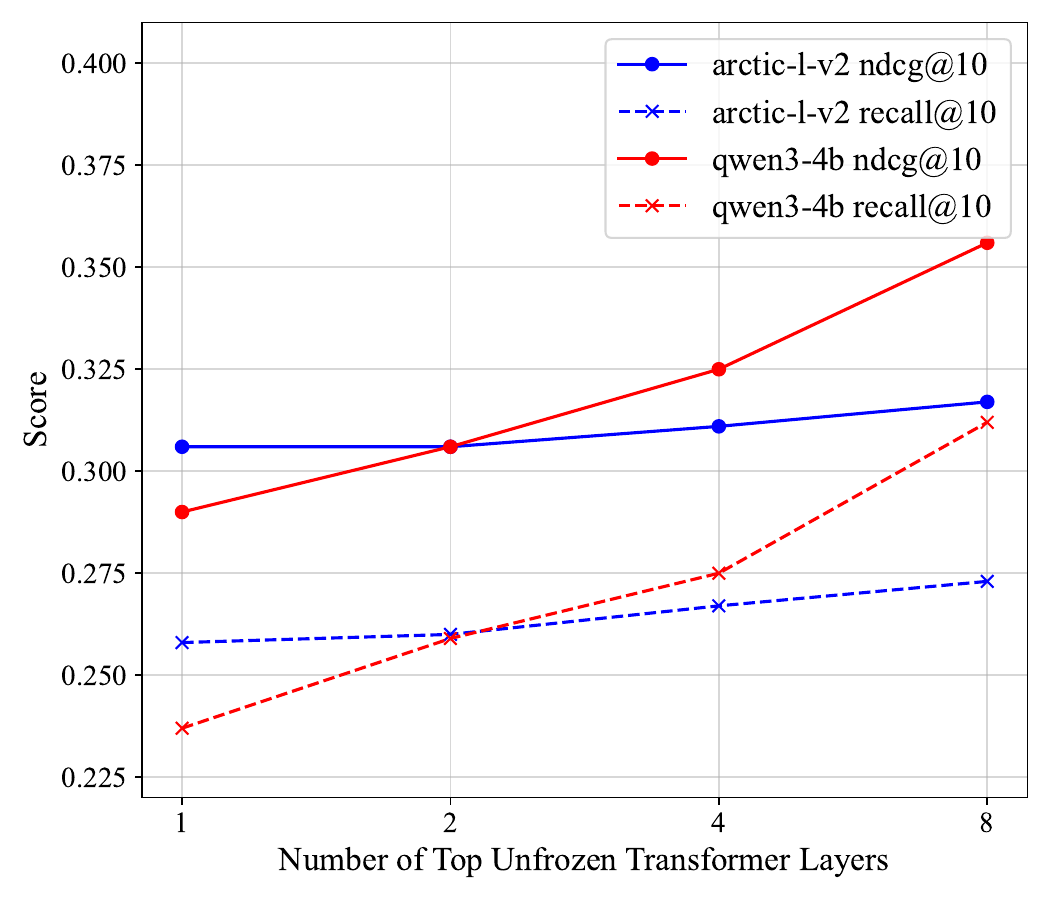}
        \caption{}
        \label{fig:c}
    \end{subfigure}

    \caption{Ablation results for query-only adaptation for DevRev-Search across (a) varying LoRA ranks (with $\alpha=r$), (b) different combinations of target LoRA modules, and (c) varying number of top unfrozen transformer layers with the rest of the model frozen}
    \label{fig:three_figs}
\end{figure*}

Standard bi-encoder fine-tuning updates both the query ($E_q$) and document ($E_d$) encoders, but modifying $E_d$ requires costly re-embedding and index reconstruction (e.g. HNSW \cite{DBLP:journals/pami/MalkovY20}) over possibly millions of documents, which is impractical in large-scale multi-tenant systems. To avoid this cost, we propose and study \textbf{Query-Only Adaptation}, which fine-tunes only the query encoder while keeping the document encoder and the index frozen. 

We evaluate on three datasets with varying relevance densities: DevRev-Search (enterprise, high density; 13.6 relevant chunks/query), SciFact (scientific, low density; 1.1), and FiQA-2018 (financial, medium density; 2.6), emphasizing recall, precise evidence retrieval, and balanced precision–recall, respectively. We use \texttt{snowflake-arctic-embed-l-v2.0} \cite{yu2024arctic} and \texttt{Qwen3-Embedding-4B} \cite{qwen3embedding} as backbone encoders. 


Training uses InfoNCE loss with mined hard negatives to improve discrimination among similar documents. A fixed tuned temperature is used, and asynchronous ANCE training \cite{xiong2020approximate} improves stability and performance. To address the limited size of DevRev-Search, we apply LLM-based augmentation, expanding the training set to 3014 queries. The open-sourced benchmark excludes these augmented queries, but they can be easily reproduced using the procedure in Appendix~\ref{app:expt_setup}.

Our analysis focuses on: (1) \textbf{Query-Only vs. Query–Document Fine-Tuning}, measuring the performance impact of freezing the document index; and (2) \textbf{Parameter-Efficient Query Adaptation},
evaluating whether low-rank and parameter-efficient methods can match full fine-tuning while enabling cost-effective multi-tenant deployment.

\subsection{Query-Only vs Query-Document Fine-Tuning}
\label{subsec:fine_tuning_strategies}

We evaluate whether asymmetric Query-Only fine-tuning ($Q$) can match symmetric Query-Document fine-tuning ($QD$). Table~\ref{tab:main_table} summarizes results across both architectures and shows that freezing the document encoder incurs minimal performance loss. In all cases, $Q$ substantially improves over the untrained baseline and remains competitive with the more expensive $QD$ approach, achieving comparable retrieval metrics across domains and, for qwen3-4b on SciFact, slightly outperforming joint tuning in Recall@10.

\subsection{Parameter Efficient Fine-Tuning}

We evaluate whether Query-Only full fine-tuning performance can be achieved with PEFT methods that reduce trainable parameters for scalable multi-tenant adaptation. We test LoRA, a linear projection head, a 2-layer feed-forward network (FFN) on embeddings, and unfreezing top 8 transformer layers. Detailed comparison of trainable parameter counts is available in Table~\ref{tab:param_efficiency} (Appendix~\ref{app:expt_setup}).

Results in Table~\ref{tab:linear_head} show that best LoRA configuration consistently matches or outperforms full fine-tuning, likely due to implicit regularization from residual learning \cite{xin-etal-2024-beyond}. Linear and FFN heads are also competitive, with FFN surpassing full fine-tuning on SciFact. In contrast, unfreezing top 8 layers underperforms despite using more parameters, highlighting LoRA’s effectiveness.

\paragraph{LoRA Rank Ablations.}
\label{subsec:rank_analysis}

We study sensitivity to the LoRA rank ($r$) with $\alpha = r$. On DevRev-Search (Fig.~\ref{fig:a}), arctic-l-v2 improves monotonically and saturates at higher ranks, while qwen3-4b peaks at $r=32$, likely due to overfitting at larger ranks. Results on SciFact and FiQA (Table~\ref{tab:lora_ablation}, Appendix~\ref{app:ablation}) show a consistent pattern: $r \in [32,64]$ provides the best trade-off between capacity and regularization. Although $r=128$ can yield minor gains, performance largely saturates, making $32$ or $64$ the most efficient choices. 

\paragraph{LoRA Target Module Ablations.}
\label{subsec:module_analysis}

We compare LoRA-based query-only adaptation across transformer sub-layer targets: dense, QV (Query-Value), QKV (Query-Key-Value), and all modules. Figure~\ref{fig:b} shows DevRev-Search results, while SciFact and FiQA results appear in Table~\ref{tab:lora_module_ablation} (Appendix~\ref{app:ablation}). While adapting all modules performs best, tuning only dense layers offers a strong performance–efficiency trade-off, especially for qwen3-4b.

\paragraph{Transformer Layer Ablations.}

We freeze the base model and vary the number of unfrozen top transformer layers. As shown in Fig.~\ref{fig:c}, performance increases monotonically with more unfrozen layers on DevRev-Search, regardless of architecture. Results on the other datasets (Table~\ref{tab:unfrozen_heads}, Appendix~\ref{app:ablation}) show the same trend.

\section{Conclusion}

 In this paper we proposed an enterprise-focused retrieval benchmark constructed via a scalable pipeline that combines multi-retriever fusion and LLM-based filtering to generate high-quality annotated data. We also proposed and systematically studied query-only fine-tuning via PEFT as a practical approach to domain adaptation of dense retrieval models for multi-tenant search, wherein only the query encoder undergoes fine-tuning, and the document index remains frozen. Our results suggest that this approach provides strong performance without the re-indexing cost, while improving the quality–efficiency trade-off.

\section*{Impact Statement}

This paper presents work whose goal is to advance the field of Machine
Learning. There are many potential societal consequences of our work, none
which we feel must be specifically highlighted here.

\nocite{langley00}

\bibliography{example_paper}
\bibliographystyle{icml2026}

\newpage
\appendix

\clearpage

\section{Appendix}
\label{app:dataset_details}

\subsection{Dataset Statistics}
\label{app:dataset_statictics}
We partition the DevRev-Search dataset, consisting of 383 high-quality queries and 65.2K chunks, into a training set and an evaluation (test) set. The dataset consists of 291 training queries and 92 test queries. The training set exhibits a rich density of relevant documents, with an average of $13.61$ golden chunks per query. Notably, the distribution of relevant documents is right-skewed, with a median of 6 and a high standard deviation ($\sigma = 21.41$). This variance reflects the diverse nature of enterprise search data, where some queries address specific technical identifiers with a single relevant source, while others address broad architectural topics with many relevant documentation fragments. To maintain the integrity of our public benchmark, we withhold the gold labels for the test queries to facilitate blind evaluation.

\subsection{Query Cleaning}
\label{app:query_cleaning}
During exploratory analysis of the collected customer queries, we observed that queries at the extremes of the length distribution were often unsuitable. Very short queries (e.g., “Hello”, “Help me please”) typically lacked sufficient context to infer user intent and were frequently associated with test or placeholder inputs. Conversely, very long queries often consisted of pasted document excerpts, logs, or code snippets, reflecting attempts to process bulk content rather than genuine customer questions. To address this, we applied percentile-based length filtering, removing queries in the bottom and top 25\% of the length distribution. Empirically, this range provided a balanced subset of queries that were more representative of natural user interactions and better suited for retrieval-based evaluation. We restricted the dataset to English-language queries as a deliberate design choice to construct a linguistically consistent benchmark. While this limits multilingual applicability, it reduces variability introduced by cross-lingual differences and enables more controlled evaluation. To prevent over-representation of repeated inputs, we removed exact duplicate queries. However, recognizing that lexical deduplication alone does not ensure semantic diversity, we further applied a clustering-based sampling strategy. Specifically, we embedded each query using the text-embedding-3-large model and performed clustering in the embedding space using DBSCAN \cite{dbscan}. Queries were then sampled in a round-robin fashion across clusters until the desired dataset size was reached, which would allow for our annotation procedure under the cost constraints we were operating with. This approach ensured coverage of diverse semantic intents while avoiding redundancy.

\subsection{Retriever Contributions}

Analysis for retriever contributions is summarized in Tables ~\ref{tab:ind_cont} and ~\ref{tab:red_cont}.
\label{app:retriever_contrib}

\begin{table}[h]
 \caption{\label{tab:ind_cont}Recall@420 of Individual Retrievers on \textbf{DevRev-Search} dataset}
  \begin{center}
  \begin{tabular}{lc}
    \hline
    \textbf{Model} & \textbf{Recall} \\
    \hline
    gemini-embedding-001     & 82.48           \\
    gte-Qwen2-7B-instruct     & 82.25           \\
    SFR-Embedding-Mistral     & 79.20           \\
    text-embedding-3-large     & 75.54           \\
    Qwen3-Embedding-8B      & 70.12            \\
    embed-english-v3     & 65.83           \\
    BM25     & 52.18           \\\hline
  \end{tabular}
  \end{center}

\end{table}

\begin{table}[h]
\caption{\label{tab:red_cont}Leave-one-out ablation study on \textbf{DevRev-Search} dataset}
  \begin{center}
  \begin{tabular}{lc}
    \hline
    \textbf{Model Combination} & \textbf{Recall} \\
    \hline
    All $\setminus$ \{gemini-embedding-001\}     & 93.25           \\
    All $\setminus$ \{gte-Qwen2-7B-instruct\}     & 95.86           \\
    All $\setminus$ \{SFR-Embedding-Mistral\}     & 96.30           \\
    All $\setminus$ \{text-embedding-3-large\}     & 97.13           \\
    All $\setminus$ \{Qwen3-Embedding-8B\}      & 96.83            \\
    All $\setminus$ \{embed-english-v3\}     & 95.61           \\
    All $\setminus$ \{BM25\}     & 95.96           \\\hline
  \end{tabular}
  \end{center}
\end{table}

\subsection{LLM-based Filtering Prompt}
\label{app:llm_filter_prompt}

We used \texttt{gpt-5-2025-08-07} model under default decoding settings of temperature 1 and top-p 1. The prompt used for LLM-based filtering consists of annotation instructions, few-shot examples, and the target query-chunk pair. The full prompt is presented below.

\subsubsection{System Prompt}
\begin{lstlisting}[breaklines=true, breakatwhitespace=true, basicstyle=\ttfamily\small]
Annotation Instructions
The focus is on whether an article chunk would help a support agent answer a query. Key instructions for annotators:

Focus on Problem in the query and Information in article chunk: Determine if the problem described in the query can be answered by the information present in article chunk. If article chunk's information would likely answer the query, then article chunk should be labeled as relevant. For example, if the article chunk explains how to use certain features in the app and the query is also asking how to use those features (even if in different words).

IMPORTANT - Beware of Superficial Word Overlap: Do not label an article chunk as relevant only because it shares some keywords with the query. Read the article chunk and query fully - article chunk and query might both mention a common term (like "login") but could be about different aspects of login (one about UI for the login page, another about authentication). Only consider lexical overlap meaningful if the article chunk contains information to answer the query (e.g. the query asks how to solve a specific login issue, and the article chunk contains information to solve that specific login issue).

{few_shot_examples}

Edge case: In the case article chunk contains only partial information required to answer the query, label it relevant only in the case when it answers the query substantially. Simple lexical overlap does not imply relevance. When in doubt, ask: "Would a support agent benefit from seeing the article chunk while answering the query?" If yes, label it similar; if not, or only minimally, then it's not relevant enough to help in the support workflow.
\end{lstlisting}

\subsubsection{Few-Shot Examples}
\begin{lstlisting}[breaklines=true, breakatwhitespace=true, basicstyle=\ttfamily\small]
Examples of Relevant article chunk:

Example 1: Query: "Where can I find the DevRev API documentation?" and article chunk: "Resources to learn how to use DevRev APIs can be found at https://developer.devrev.ai/". The query is asking about where to find documentation on how to use DevRev APIs and the article chunk contains the information about the location where to find DevRev API documentation. The article chunk should be marked as relevant - it contains the information required to answer the query (even if words differ).

Example 2: Query: "What is a custom object?" and article chunk: "To create a custom object raise a support ticket. Custom objects are DevRev objects which can be customized". Even though the article chunk initially contains the information on how to create a custom object, it later also contains the information on what is a custom object which is what is asked in the query. Mark the article chunk relevant.

Examples of Non-Relevant article chunk:

Example 1: Query: "How to create a vista?" vs article chunk: "Vista is a list of DevRev objects" Both query and article chunk are about vistas and share the word "vista" but the information in article chunk is different from what query is asking about (query is asking how to create a vista, the article chunk is about what are vistas). This article chunk should be marked non relevant - information in the article chunk would not help answer the query.

Example 2: Query: "How to solve FORBIDDEN error when calling custom object API?" vs article chunk: "To solve BAD_REQUEST error when calling custom object API, look for DevRev custom object API documentation and fix your request structure". On the surface the article chunk looks relevant (same feature: custom object API). However, the error nature is different (one is a FORBIDDEN error, another is a BAD_REQUEST error). Unless further context in the article chunk reveals that both are the same errors, treat the article chunk as non relevant because the resolutions of both errors would differ (one might need more permissions, the other requires fixing the request).

Example 3: Query: Resource Center downloads tutorials API documentation vs article chunk: "Prerequisites\n* Send your first API request\n* Making a GET request\n* Next steps\n\nAPI Reference\n\nGetting started\n===============\n\nCopy page\n\nThe DevRev API is organized around REST. Our API has predictable resource-oriented URLs, accepts". On the surface the article chunk appears to be answering the query because it has links to the documentation but the problem is that it is part of start of a webpage so only has relative links and not actual content or complete URL.
\end{lstlisting}

\subsubsection{User Prompt Template}
\begin{lstlisting}[breaklines=true, breakatwhitespace=true, basicstyle=\ttfamily\small]
Query: {query}
Article Chunk: {candidate}
\end{lstlisting}




\subsection{Motivation for the Multi-Tenant Deployment Setting}
\label{app:multi_tenant}

In our multi-tenant scenario, each tenant maintains its own document corpus and independent document index. No index is shared across tenants, ensuring strict privacy and isolation. Multi-tenancy is achieved through a shared base dense retriever combined with tenant-specific query adapters implemented as lightweight LoRA modules, or other PEFT adapters. Each tenant trains its adapter using only its own queries, and at serving time, incoming queries are routed to the appropriate tenant so that the corresponding adapter is dynamically activated while the base model remains shared. This design keeps both indices and adapters fully isolated while enabling efficient multi-tenant retrieval.

This approach is motivated by practical considerations. Classical fine-tuning would require updating both query and document encoders and re-indexing each tenant’s corpus—an expensive and operationally heavy process. By adapting only the query side, document embeddings and indices remain fixed, avoiding re-indexing. In our experiments, we simulate heterogeneous tenants using three diverse domains (Enterprise, Scientific, Financial) with different relevance distributions. Tenant-specific adapters are lightweight and inexpensive to store and manage, demonstrating the feasibility of this deployment approach in practice.

\subsection{Experimental Setup}
\label{app:expt_setup}

All our training optimizes InfoNCE loss using AdamW optimizer, with 16 mined hard negatives per query, with 8 negatives randomly sampled from them each time. A learning rate of $5 \times 10^{-6}$ is used for all runs with a cosine scheduler and no warm-up. For all the experiments, we use a uniform temperature of 0.1, which was found to work the best.

\begin{figure}[H]
    \centering
    \includegraphics[width=0.5\textwidth]{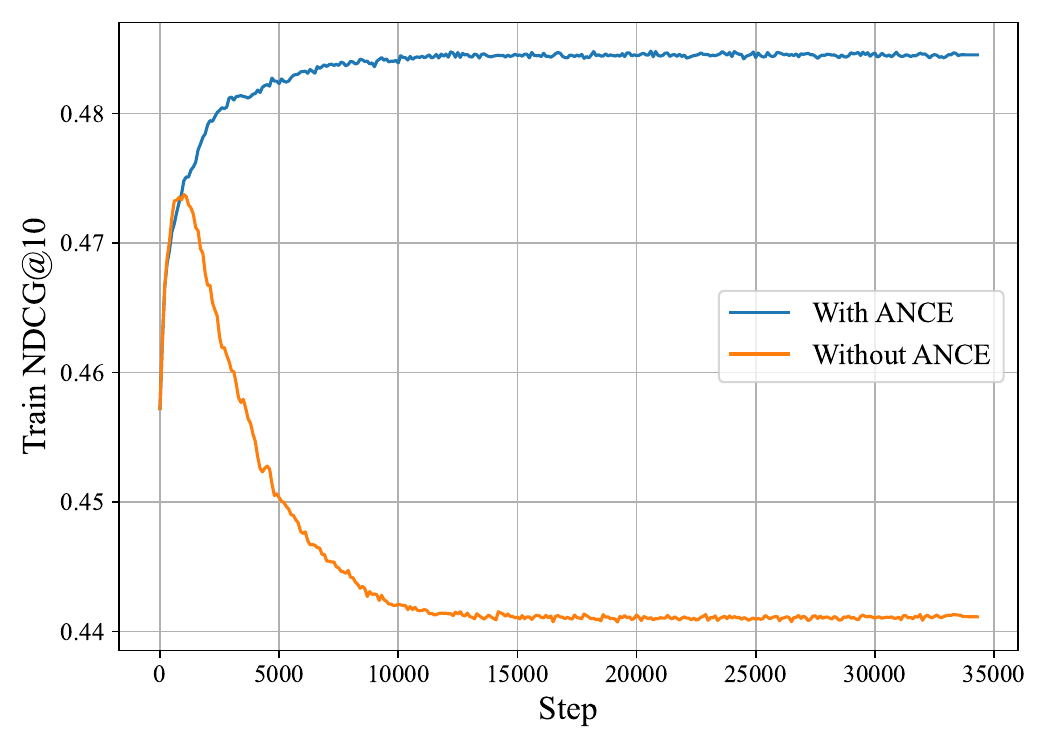}
    \caption{Training NDCG@10 vs training step for Query-Only adaptation of arctic-l-v2 (full model) }
    \label{fig:training_curve}
\end{figure}

Additionally, we noticed that simply using fixed mined hard negatives leads to cases of representation collapse wherein InfoNCE loss keeps improving, but training metrics rapidly deteriorate after a point, especially in datasets like FiQA and DevRev-Search. Upon further analysis, we noticed that this happens due to evolving hard negatives, where over the course of training, old hard negatives become 'simple' and new hard negatives replace them. To counteract this, we adopt the asynchronous hard negative refresh technique proposed by ANCE \cite{xiong2020approximate} to periodically update the hard negatives in InfoNCE loss at every 200 steps. This proved to improve the performance in most cases, in addition to much more stable training as shown in Fig \ref{fig:training_curve}. Quantitative comparison of results obtained with and without this technique is shown in Table \ref{tab:ance_table}

\textbf{Training Set Augmentation:} For augmenting the training set of DevRev-Search, we used a carefully prompted LLM to take in an individual query as well as their ground truth chunks, and generate a set of new queries, where each of the generated query addresses concepts/topics/questions that are subsets of the concepts/topics/questions addressed in the original query from which it was created. This achieves three goals : (1) the ground truth chunks pertaining to the generated query would be a subset of the ground truth chunks of the original query, (2) the generated queries follow the same ‘distribution’ as the original queries, (3) the generated queries are strictly semantically distinct from the original query so as to ensure no data duplication. Applying this strategy to the original 291 train queries resulted in a total of 3014 queries. Following this, for each of the generated queries, we filter out the irrelevant chunks from the ground truth chunks of the original query using an LLM, which we then take as the ground truth chunks for the generated query. This resulted in a diverse augmented set that pertained to the same ‘distribution’ and, at the same time, mitigated risks that came from a small training set. 

The feed-forward network (FFN) used in our PEFT experiments consists of GELU \cite{hendrycks2016gaussian} activations and 2 hidden layers, each with the same number of nodes as the base model's embedding dimension. Also, for both linear and FFN heads, we find that zeroing biases and initializing weight matrices to the identity perform consistently better than standard initializations such as Xavier \cite{glorot2010understanding} or Kaiming \cite{he2015delving}.

\subsection{Ablation Results}
\label{app:ablation}
Tables \ref{tab:lora_ablation}, \ref{tab:lora_module_ablation}, and \ref{tab:unfrozen_heads} present the ablation results across different LoRA ranks, different LoRA target modules, and different numbers of unfrozen transformer layers of the dense model, for Query-Only fine-tuning, respectively.

\clearpage

\begin{table*}[t]
\caption{\label{tab:ance_table}Retrieval effectiveness with and without ANCE for full Query-Only adaptation}
\begin{center}
\begin{tabular}{llcccccc}
\toprule
\multirow{2}{*}{Model} & \multirow{2}{*}{} 
& \multicolumn{2}{c}{DevRev-Search} 
& \multicolumn{2}{c}{SciFact} 
& \multicolumn{2}{c}{FiQA} \\
\cmidrule(lr){3-4} \cmidrule(lr){5-6} \cmidrule(lr){7-8}
& & recall@10 & ndcg@10 & recall@10 & ndcg@10 & recall@10 & ndcg@10 \\
\midrule

\multirow{2}{*}{arctic-l-v2}
& Standard    & 0.291 & 0.331 & \textbf{0.860} & \textbf{0.731} & 0.543 & 0.473 \\
& +ANCE   & \textbf{0.296} & \textbf{0.343} & 0.854 & 0.727 & \textbf{0.554} & \textbf{0.478} \\

\midrule

\multirow{2}{*}{qwen3-4b}
& Standard    & \textbf{0.333} & \textbf{0.373} & 0.941 & 0.813 & 0.679 & 0.593 \\
& +ANCE   & 0.327 & 0.370 & \textbf{0.953} & \textbf{0.830} & \textbf{0.688} & \textbf{0.602} \\

\bottomrule
\end{tabular}

\end{center}
\end{table*}

\begin{table*}[t]
\caption{Parameter efficiency comparison for Query-Only adaptation across LoRA ranks, unfrozen layers, and head-only training.}
\begin{center}
\label{tab:param_efficiency}
\begin{tabular}{llrrr}
\toprule
\multirow{2}{*}{Model} & \multirow{2}{*}{Configuration} & \multicolumn{1}{c}{Trainable} & \multicolumn{1}{c}{Total} & \multicolumn{1}{c}{\% Full} \\
 &  & \multicolumn{1}{c}{Params} & \multicolumn{1}{c}{Params} & \\
\midrule

\multirow{12}{*}{arctic-l-v2}
& Full Model                 &   566,705,152 &   566,705,152 & 100.00 \\
& LoRA $r=8$                   &     3,538,944 &   570,244,096 &   0.62 \\
& LoRA $r=16$                  &     7,077,888 &   573,783,040 &   1.25 \\
& LoRA $r=32$                  &    14,155,776 &   580,860,928 &   2.50 \\
& LoRA $r=64$                  &    28,311,552 &   595,016,704 &   4.99 \\
& LoRA $r=128$                 &    56,623,104 &   623,328,256 &   9.99 \\
& Linear Head (Frozen Base)  &     1,049,600 &   567,754,752 &   0.19 \\
& MLP Head (Frozen Base)     &     3,148,800 &   569,853,952 &   0.56 \\
& 1 Unfrozen Layer           &    12,596,224 &   566,705,152 &   2.22 \\
& 2 Unfrozen Layers          &    25,192,448 &   566,705,152 &   4.45 \\
& 4 Unfrozen Layers          &    50,384,896 &   566,705,152 &   8.89 \\
& 8 Unfrozen Layers          &   100,769,792 &   566,705,152 &  17.78 \\

\midrule

\multirow{12}{*}{qwen3-4b}
& Full Model                 & 4,021,774,336 & 4,021,774,336 & 100.00 \\
& LoRA $r=8$                   &    16,515,072 & 4,038,289,408 &   0.41 \\
& LoRA $r=16$                  &    33,030,144 & 4,054,804,480 &   0.82 \\
& LoRA $r=32$                  &    66,060,288 & 4,087,834,624 &   1.64 \\
& LoRA $r=64$                  &   132,120,576 & 4,153,894,912 &   3.29 \\
& LoRA $r=128$                 &   264,241,152 & 4,286,015,488 &   6.57 \\
& Linear Head (Frozen Base)  &     6,556,160 & 4,028,330,496 &   0.16 \\
& MLP Head (Frozen Base)     &    19,668,480 & 4,041,442,816 &   0.49 \\
& 1 Unfrozen Layer           &   100,933,376 & 4,021,774,336 &   2.51 \\
& 2 Unfrozen Layers          &   201,864,192 & 4,021,774,336 &   5.02 \\
& 4 Unfrozen Layers          &   403,725,824 & 4,021,774,336 &  10.04 \\
& 8 Unfrozen Layers          &   807,449,088 & 4,021,774,336 &  20.08 \\

\bottomrule
\end{tabular}
\end{center}
\end{table*}

\clearpage

\begin{table*}[t]
\caption{\label{tab:lora_ablation} Retrieval effectiveness variation across different LoRA ranks ($r=\alpha$ for all configurations)}
\begin{center}
\begin{tabular}{llcccccc}
\toprule
\multirow{2}{*}{Model} & \multirow{2}{*}{\makecell{LoRA\\Rank}} 
& \multicolumn{2}{c}{DevRev-Search} 
& \multicolumn{2}{c}{SciFact} 
& \multicolumn{2}{c}{FiQA} \\
\cmidrule(lr){3-4} \cmidrule(lr){5-6} \cmidrule(lr){7-8}
& & recall@10 & ndcg@10 & recall@10 & ndcg@10 & recall@10 & ndcg@10 \\
\midrule

\multirow{4}{*}{arctic-l-v2}
& 8    & 0.258 & 0.307 & 0.867 & 0.738 & 0.540 & 0.469 \\
& 16   & 0.268 & 0.322 & 0.880 & 0.748 & 0.548 & 0.474 \\
& 32   & 0.283 & 0.328 & 0.905 & 0.769 & 0.554 & \textbf{0.478} \\
& 64   & 0.304 & 0.341 & 0.914 & 0.807 & \textbf{0.555} & \textbf{0.478} \\
& 128   & \textbf{0.309} & \textbf{0.342} & \textbf{0.915} & \textbf{0.827} & \textbf{0.555} & \textbf{0.478} \\

\midrule

\multirow{4}{*}{qwen3-4b}
& 8    & 0.304 & 0.346 & 0.948 & 0.826 & 0.685 & 0.603 \\
& 16   & 0.329 & 0.360 & 0.948 & 0.832 & 0.686 & 0.605 \\
& 32   & \textbf{0.355} & \textbf{0.399} & 0.950 & \textbf{0.844} & 0.690 & \textbf{0.607} \\
& 64   & 0.336 & 0.369 & 0.950 & 0.816 & \textbf{0.694} & 0.607 \\
& 128   & 0.336 & 0.380 & \textbf{0.957} & 0.831 & 0.682 & 0.601 \\

\bottomrule
\end{tabular}
\end{center}
\end{table*}
\begin{table*}[t]
\caption{\label{tab:lora_module_ablation}Retrieval effectiveness variation across different target modules for LoRA $r=64$ and $\alpha=64$}
\begin{center}
\begin{tabular}{llcccccc}
\toprule
\multirow{2}{*}{Model} & \multirow{2}{*}{Layers} 
& \multicolumn{2}{c}{DevRev-Search} 
& \multicolumn{2}{c}{SciFact} 
& \multicolumn{2}{c}{FiQA} \\
\cmidrule(lr){3-4} \cmidrule(lr){5-6} \cmidrule(lr){7-8}
& & recall@10 & ndcg@10 & recall@10 & ndcg@10 & recall@10 & ndcg@10 \\
\midrule

\multirow{4}{*}{arctic-l-v2}
& Dense    & 0.287 & 0.329 & 0.854 & 0.726 & 0.551 & 0.477 \\
& QV   & 0.283 & 0.325 & 0.854 & 0.726 & 0.554 & 0.476 \\
& QKV   & 0.283 & 0.325 & 0.854 & 0.726 & 0.554 & 0.477 \\
& All   & \textbf{0.304} & \textbf{0.341} & \textbf{0.914} & \textbf{0.807} & \textbf{0.555} & \textbf{0.478} \\

\midrule

\multirow{4}{*}{qwen3-4b}
& Dense    & 0.333 & 0.371 & \textbf{0.950} & \textbf{0.818} & 0.693 & \textbf{0.607} \\
& QV   & 0.324 & \textbf{0.377} & 0.923 & 0.793 & 0.688 & 0.605 \\
& QKV   & \textbf{0.337} & 0.368 & 0.930 & 0.796 & 0.689 & 0.606 \\
& All   & 0.336 & 0.369 & \textbf{0.950} & 0.816 & \textbf{0.694} & \textbf{0.607} \\

\bottomrule
\end{tabular}
\end{center}
\end{table*}

\begin{table*}[t]
\caption{\label{tab:unfrozen_heads}Retrieval effectiveness variation across different numbers of unfrozen transformer layers at the top of the base model, with the rest of the layers frozen}
\begin{center}
\begin{tabular}{llcccccc}
\toprule
\multirow{2}{*}{Model} & \multirow{2}{*}{Layers} 
& \multicolumn{2}{c}{DevRev-Search} 
& \multicolumn{2}{c}{SciFact} 
& \multicolumn{2}{c}{FiQA} \\
\cmidrule(lr){3-4} \cmidrule(lr){5-6} \cmidrule(lr){7-8}
& & recall@10 & ndcg@10 & recall@10 & ndcg@10 & recall@10 & ndcg@10 \\
\midrule

\multirow{4}{*}{arctic-l-v2}
& 1    & 0.258 & 0.306 & 0.827 & 0.700 & 0.534 & 0.462 \\
& 2   & 0.260 & 0.306 & 0.835 & 0.709 & 0.536 & 0.464 \\
& 4   & 0.267 & 0.311 & 0.835 & 0.712 & 0.539 & 0.471 \\
& 8   & \textbf{0.273} & \textbf{0.317} & \textbf{0.849} & \textbf{0.718} & \textbf{0.547} & \textbf{0.476} \\

\midrule

\multirow{4}{*}{qwen3-4b}
& 1    & 0.237 & 0.290 & 0.903 & 0.778 & 0.661 & 0.576 \\
& 2   & 0.259 & 0.306 & 0.908 & 0.783 & 0.664 & 0.579 \\
& 4   & 0.275 & 0.325 & 0.912 & 0.788 & 0.667 & 0.583 \\
& 8   & \textbf{0.312} & \textbf{0.356} & \textbf{0.932} & \textbf{0.802} & \textbf{0.673} & \textbf{0.590} \\

\bottomrule
\end{tabular}
\end{center}
\end{table*}


\end{document}